\documentclass[twocolumn,prb,showpacs]{revtex4}
\usepackage{graphicx}
\usepackage{ulem}

\newcommand{\de}{\partial}

\begin{document}

\title{Temperature-dependent resistivity of ferromagnetic GaMnAs: Interplay between impurity scattering and many-body
effects}

\author{F. V. Kyrychenko}
\author{C. A. Ullrich}
\affiliation{Department of Physics and Astronomy, University of
Missouri, Columbia, Missouri 65211, USA}

\date{\today}

\begin{abstract}
The static conductivity of the dilute magnetic semiconductor
Ga$_{1-x}$Mn$_x$As is calculated using the memory function
formalism and time-dependent density-functional theory to account
for impurity scattering and to treat Hartree and exchange
interactions within the hole gas. We find that the Coulomb
scattering off the charged impurities alone is not sufficient to
explain the experimentally observed drop in resistivity below the
ferromagnetic transition temperature: the often overlooked
scattering off the fluctuations of localized spins is shown to
play a significant role.
\end{abstract}


\maketitle

The perspective of utilizing the charge {\it and} the spin of the
electrons for new electronic device applicati\-ons has generated
tremendous interest in the field of spintronics. \cite{spintronic}
A unique combination of magnetic and semiconducting properties
makes dilute magnetic semiconductors (DMSs) very attractive for
various spintronics applications. \cite{ohno} Among the family of
DMSs, much attention has been paid to Ga$_{1-x}$Mn$_x$As since the
discovery of its relatively high ferromagnetic transition
temperature, \cite{ohno} with a current record of $T_c = 185$ K.
\cite{record}

Unlike most other III-V DMSs, the nature of the itinerant carriers
in Ga$_{1-x}$Mn$_x$As is still under debate. It is widely accepted
that for low-doped insulating samples the Fermi energy lies in a
narrow impurity band. For more heavily doped, high-$T_c$ metallic
samples there are strong indications that the impurity band merges
with the host semiconductor valence band forming mostly host-like
states at the Fermi energy with some low-energy tail of
disorder-related localized states. \cite{jungwirth07PRBtails}
First-principles calculations \cite{mahadevan,sandratskii,yildrim}
have so far not been fully conclusive regarding the nature of the
itinerant carriers in this case, and further theoretical studies
continue to be necessary. The question is thus, in essence,
whether the valence band \cite{dietl01} or impurity band
\cite{berciu} picture is more adequate to describe the various
experimental results in Ga$_{1-x}$Mn$_x$As.

The purpose of this paper is to present a study which supports the
valence band picture for electronic transport properties and for
the optical conductivity in Ga$_{1-x}$Mn$_x$As. In this material,
unlike in II-VI DMSs, the magnetic ions in substitutional
positions act as acceptors delivering holes and producing not only
localized spins but also charged defects. We will argue that it is
important to treat disorder and many-body effects beyond the
simple relaxation time and static screening models which were used
in earlier theoretical studies.\cite{sinova,rates,lopez,ewelina}

Lopez-Sancho and Brey \cite{lopez} considered the temperature
dependence of the Coulomb scattering off the acceptor centers and
found that the carrier relaxation rate is reduced by around 20\%
in the ferromagnetic phase, consistent with the experimentally
observed drop in resistivity. \cite{potashnik,edmonds,matsukura} This drop was
attributed entirely to the effects associated with the scattering
off Coulomb disorder. The main mechanism found to be responsible
for the observed drop in resistivity was the change of the
semiconductor band structure: during the transition from the
paramagnetic to the ferromagnetic state, the giant spin splitting
of the energy bands substantially modifies the shape of the Fermi
surface, thus altering the possible scattering wave vectors and,
consequently, the magnitude of scattering matrix elements.

These findings speak in favor of the valence band picture of
Ga$_{1-x}$Mn$_x$As. However, the model of Ref.~\onlinecite{lopez}
employed a simplified treatment for the screening of the charge
disorder by itinerant carriers, neglecting the exchange part of the
electron-electron interaction within the hole liquid. Furthermore,
the scattering off the fluctuations of localized spins was ignored.
We will show that both effects play an important role in spin
polarized systems and should be included in the valence band picture
model of the itinerant holes in Ga$_{1-x}$Mn$_x$As. In fact, our
calculations suggest that the previously suggested
origin\cite{lopez} of the resistivity drop in the ferromagnetic
phase should be revised: the main reduction of the scattering rate
comes from the suppression of the fluctuations of localized spins in
the magnetically ordered state.

Earlier we  developed a theory of transport in charge and spin
disordered media which combines a multiband ${\bf k\cdot p}$
approach for an accurate description of the valence band states
with a more comprehensive treatment of disorder and
electron-electron interaction.\cite{KUPRB07,KUjpcm09} Our theory
is based on an equation of motion approach for the current-current
response function \cite{Gotze72,GV} and has some similarity with
the memory function formalism. \cite{Gotze81,belitz,UV}

In this paper we apply our formalism to describe the transport
properties of Ga$_{1-x}$Mn$_x$As in the static regime. Specifically,
we focus on the pronounced drop in resistivity below $T_c$ which has
been observed for optimally annealed metallic samples.
\cite{potashnik} To investigate this phenomenon let us first look at
the standard expression for the static conductivity obtained from
the semiclassical Boltzmann equation: \cite{rates}
\begin{equation} \label{stat_cond}
\sigma_{\alpha \beta}=\frac{e^2}{\hbar V}\sum_{n,{\bf k}}
\frac{\tau_{n,{\bf k}}}{\hbar} \frac{\de E_{n,{\bf k}}}{\de
k_{\alpha}} \frac{\de E_{n,{\bf k}}}{\de k_{\beta}}
\delta(E_{n,{\bf k}}-E_f),
\end{equation}
where the summation is over the wave vector ${\bf k}$ and the energy
band index $n$. The part of Eq.~(\ref{stat_cond}) which is most
sensitive to temperature is the carrier scattering rate
$\tau^{-1}_{n,{\bf k}}$. The task is therefore to derive a
microscopic expression for $\tau^{-1}_{n,{\bf k}}$ which accounts
for all relevant scattering mechanisms, as well as for electronic
many-body effects.

Our model for the itinerant carriers in Ga$_{1-x}$Mn$_x$As is that
of a system with charge and spin disorder described by the
Hamiltonian
\begin{equation}\label{HI4}
  \hat{H}_I = V^2 \sum_{\bf k} \hat{\vec{\cal{U}}}({\bf k})\cdot \hat{\vec{\rho}}({\bf -k}),
\end{equation}
where the four-component disorder scattering potential
\begin{equation}\label{pot}
  \hat{\vec{\cal{U}}}({\bf k})=\frac1{V}\sum_j \left(\begin{array}{c}
    U_j({\bf k}) \\
    \frac{J}2 \hat{S}_j^- \\
    \frac{J}2 \hat{S}_j^+ \\
    \frac{J}2\left( \hat{S}_j^z-\langle S \rangle \right) \
  \end{array} \right) e^{i{\bf k\cdot R}_j}
\end{equation}
is coupled to the four-component charge and spin density operator
of the band carriers,
\begin{equation}\label{rho}
  \hat{\rho}^{\mu}({\bf k})=\frac1{V}\sum_{\bf q} \sum_{n n'} \langle u_{n',{\bf q-k}}|\sigma^{\mu}| u_{n,{\bf q}}\rangle \;
  \hat{a}^+_{n',{\bf q-k}} \, \hat{a}_{n,{\bf q}} \:.
\end{equation}
Here, $\sigma^\mu$ ($\mu=1,+,-,z$) is defined via the Pauli
matrices, where $\sigma^1$ is the $2\times 2$ unit matrix,
$\sigma^\pm=(\sigma^x \pm i \sigma^y)/2$, and $| u_{n,{\bf
q}}\rangle$ are the two-component Bloch function spinors with wave
vector ${\bf q}$ and band index $n$. The summation in
Eq.~(\ref{pot}) is performed over all defects.

The general case of multiple types of defects, including defect
correlations, was considered in Ref.~\onlinecite{KUPRB07}. For
simplicity we here include only the most important defect type,
namely randomly distributed manganese ions in gallium substitutional
positions (${\rm Mn_{Ga}}$). Our model treats localized spins as
quantum mechanical operators coupled to the band carriers via a
contact Heisenberg interaction featuring a momentum-independent
exchange constant $J$. We use the value of $VJ=55\:{\rm meV\,nm^3}$,
which corresponds to the widely used DMS $p-d$ exchange constant
$N_0 \beta=1.2\,$eV. \cite{dietl01} The $z$-axis is chosen along the
direction of the macroscopic magnetization.

We obtain the following expression for the tensor of Drude-like
frequency and momentum dependent relaxation rates:
\begin{eqnarray}\label{finalS}
\tau^{-1}_{\alpha\beta}({\bf q},\omega)  & = & i \frac{V^2}{n m
\omega}\sum_{ \bf k \atop \mu\nu } k_{\alpha} k_{\beta}
  \left\langle\hat{\cal{U}}_{\mu}({\bf -k})\;
  \hat{\cal{U}}_{\nu}({\bf k})\right\rangle_{H_m} \\ \nonumber
&\times&
  \Big(\chi_{\rho^{\mu}\rho^{\nu}}({\bf q-k},\omega)-\chi^c_{\rho^{\mu} \rho^{\nu}}({\bf k},0) \Big)+ \tau^{-1}_A,
\end{eqnarray}
where $\alpha,\beta=x,y,z$ are Cartesian coordinates, $n$ is the
carrier concentration, and $\chi_{\rho^{\mu} \rho^{\nu}}({\bf
k},\omega)$ are charge- and spin-density response functions
associated with the operators (\ref{rho}). The superscript $c$ in
Eq.~(\ref{finalS}) refers to a clean (defect-free) system, and
$\tau^{-1}_A$ stands for those additional contributions which arise
in magnetically ordered systems only. Here, we just give the
clean-system and ${\bf q}=0$ limit, which suffices to illustrate the
general structure:
\begin{eqnarray}\label{tauA}
\tau^{-1}_A&=& i \frac{V^2}{n m \omega}\sum_{ \bf k \atop \mu\nu }
k_{\alpha} k_{\beta}
  \left\langle\left[\hat{\cal{U}}_{\mu}({\bf k}),\;
  \hat{\cal{U}}_{\nu}({\bf -k})\right]\right\rangle_{H_m}  \int_0^{\infty} d\tau \nonumber\\
&\times&  \frac{i V}{\hbar} \left\langle \hat{\rho}^{\nu}({\bf
-k})\; \hat{\rho}^{\mu}({\bf k},\tau)\right\rangle_{H_c} \left(
e^{i\omega\tau} - 1\right)e^{-\eta \tau} .
\end{eqnarray}
The complete expression for $\tau^{-1}_A$ as well as details of the derivation of Eq.~(\ref{finalS}) can be found in
Ref.~\onlinecite{KUjpcm09}.

All information about the itinerant carriers, including band
structure and electron-electron interaction, is contained within the
set of charge- and spin-density response functions
$\chi_{\rho^{\mu}\rho^{\nu}}({\bf k},\omega)$. Strictly speaking,
these response functions correspond to the disordered system, and
Eq.~(\ref{finalS}) should be calculated
self-consistently.\cite{Gold} We here assume that the disorder is
sufficiently weak  so that we can approximate (\ref{finalS}) by
expanding to second order in the disorder potential
$\hat{\cal{U}}({\bf k})$, and thus replace
$\chi_{\rho^{\mu}\rho^{\nu}}({\bf k},\omega)$ by its clean system
counterpart $\chi^c_{\rho^{\mu}\rho^{\nu}}({\bf k},\omega)$.

To account for the complexity of the band structure, we use a
standard 8-band ${\bf k\cdot p}$ approach with contributions from
the remote bands taken up to the second order in the wave vector.
\cite{pigeon} The mean-field part of the $p-d$ exchange
interaction between itinerant holes and localized spins causes a
spin splitting of the bands of the semiconductor host material.
Technical details of this multiband linear-response approach will
be published elsewhere.

The major advantage of Eq. (\ref{finalS}) is that it can be combined
in a straightforward manner with time-dependent density functional
theory (TDDFT),\cite{rungegross} which allows us to describe
electron-electron interaction effects, including correlations and
collective modes, in principle exactly. In TDDFT the charge- and
spin-density response functions of the interacting system are
written as follows:\cite{grosskohn}
\begin{equation}\label{tddft}
  \uuline{\chi}^{-1}({\bf q},\omega)=\uuline{\chi_0}^{-1}({\bf q},\omega) - \uuline{v}(q) - \uuline{f_{xc}} ({\bf q},\omega),
\end{equation}
where $\uuline{\chi_0}$ denotes the matrix of response functions
of the noninteracting system, $\uuline{v}(q)$ is the Hartree part
of the electron-electron interactions, and $\uuline{f_{xc}}$ is
the matrix of exchange and correlation kernels. All quantities in
Eq. (\ref{tddft}) are $4\times 4$ matrices; according to Eq.
(\ref{pot}), the first component is charge, and the other
components are spin $+$, $-$, and $z$. As a simplification we use
only the exchange part of $\uuline{f_{xc}}$ and apply the
adiabatic local spin density approximation. The local field
factors for partially spin polarized system were calculated
according to Ref.~\onlinecite{UllrichFlatte02}.

Since the mean-field part of the $p-d$ exchange interaction is
extracted from the disorder Hamiltonian, the total relaxation rate
(\ref{finalS}) can be separated into contributions associated with
Coulomb disorder and with fluctuations of the localized spins. The
transverse component (perpendicular to the magnetization) of the
relaxation rate tensor in the long wave length (${\bf q}=0$) and
static ($\omega \to 0$) limit then has the form
\begin{equation}\label{tau}
 \tau^{-1}_{xx}=\tau^{-1}_c+\tau^{-1}_s,
\end{equation}
where the charge disorder contribution is
\begin{equation}\label{tauc}
\tau^{-1}_c  = i\frac{n_i}n \frac{V}m \lim_{\omega \to 0}
\frac1{\omega} \sum_{ \bf k} k^2_x  |U(k)|^2 \Big(\chi_{nn}({\bf
k},\omega)-\chi^c_{nn}({\bf k},0) \Big)
\end{equation}
and the contribution from the fluctuations of the localized spins
is given by
\begin{eqnarray}\label{taus}
\tau^{-1}_s  & = & i\frac{n_i}n \frac{V}m \frac{J^2}4 \lim_{\omega \to 0} \frac1{\omega} \sum_{ \bf k} k^2_x \\
  &\times& \bigg[\Big(\langle \hat{S}_z^2 \rangle - \langle \hat{S}_z \rangle^2 \Big)
\Big(\chi_{s^z s^z}({\bf k},\omega)-\chi_{s^z s^z}({\bf k},0)\Big) \nonumber \\
  &+&\langle \hat{S}^- \hat{S}^+ \rangle \Big(\chi_{s^+s^-}({\bf k},\omega)-\chi_{s^+s^-}({\bf k},0)\Big) \nonumber\\
  &+& \langle \hat{S}^+ \hat{S}^- \rangle \Big(\chi_{s^-s^+}({\bf k},\omega)-\chi_{s^-s^+}({\bf k},0)\Big) \bigg]+\tau^{-1}_A. \nonumber
\end{eqnarray}
Here, $n_i$ denotes the concentration of $\rm Mn_{Ga}$ defects.
$U(k)$ represents the Coulomb potential of a single acceptor
center screened by the dielectric constant of the host material,
where we take $\varepsilon=13$ for GaAs; the screening by the
electron liquid is absorbed in the band-carrier response
functions. The angular brackets in Eq.~(\ref{taus}) denote the
thermodynamic average with respect to the magnetic subsystem
Hamiltonian $\hat{H}_m$ in Eq.~(\ref{finalS}). We assume
$\hat{H}_m$ to be a pairwise Heisenberg-like Hamiltonian. In our
calculations we use the experimental value of $T_c$ as an input
parameter and apply the standard mean field approach to obtain the
temperature dependence of thermodynamically averaged quantities in
Eq.~(\ref{taus}).

\begin{figure}
\centering
\includegraphics[width=1.0\linewidth]{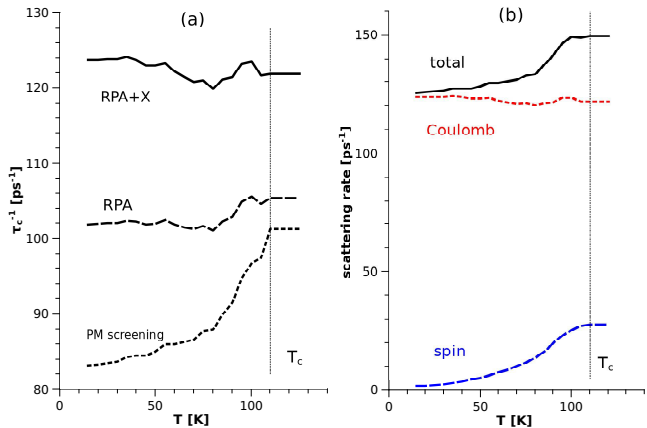}
\caption{Temperature dependence of the carrier relaxation rates in
$\rm Ga_{0.95}Mn_{0.05} As$. The vertical line indicates $T_c$. (a)
relaxation rates associated with the scattering off Coulomb
disorder, calculated within different screening models (see text for
details). (b) total relaxation rate (solid line) and contributions
from scattering off Coulomb disorder (dotted line) and off localized
spin fluctuations (dashed line).} \label{fig1}
\end{figure}

Fig.~\ref{fig1} presents the typical temperature dependence of the
carrier relaxation rate in Ga$_{1-x}$Mn$_x$As obtained within our
model. The calculations were performed for a system with Mn
concentration $x=0.05$ and carrier concentration of $p=0.3$ holes
per Mn in Ga substitutional positions. The left panel shows the
contributions from the scattering off Coulomb disorder with
screening effects accounted for by three different methods. The
dotted line corresponds to screening described within the
Thomas-Fermi approximation for paramagnetic systems. The screening
here is temperature independent and the 20\% drop in the scattering
rate in the ferromagnetic phase is entirely due to modification of
the possible scattering wave vectors, the mechanism described in
Ref~\onlinecite{lopez}.

If, however, we allow the change of the band structure to affect the
screening as well, e.g. on the random-phase approximation (RPA)
level (first two terms in Eq.~(\ref{tddft}) and dashed line in
Fig.~\ref{fig1}a), the drop in the resistivity in the ferromagnetic
phase is significantly reduced. This effect was also considered in
Ref.~\onlinecite{lopez}. But if we now go further and include the
exchange part of electron-electron interaction in Eq.~(\ref{tddft}),
then the drop in the resistivity is completely washed out, see the
solid line in Fig.~\ref{fig1}a. Moreover, for some parameters, the
trend is reversed and the scattering off Coulomb disorder actually
{\it increases} in the ferromagnetic phase.

The explanation for this behavior lies in the exchange part, which
counteracts the larger Hartree part of the electron-electron
interaction. It reduces, therefore, the screening of the Coulomb
disorder potential calculated within RPA. Therefore we have an
overall increase in the charge relaxation rate once the exchange
part of electron-electron interaction is taken into account, see
Fig.~\ref{fig1}a. On the other hand, the exchange part of the
electron-electron interaction is more pronounced for spin polarized
systems, resulting in a stronger reduction of the screening of the
disorder potential and thus causing an increase of the Coulomb
scattering in the ferromagnetic phase. This process, neglected in
Ref.~\onlinecite{lopez}, competes with and, for some parameters,
reverses the reduction of the relaxation rate due to band-structure
related modifications of the scattering wave vectors.

It is thus apparent that the scattering off the Coulomb disorder
potential alone cannot be responsible for the experimentally
observed drop in resistivity. The other possible contribution is the
scattering off the fluctuations of localized spins. In
Fig.~\ref{fig1}b we plot the temperature dependence of the
scattering rate for both mechanisms. The scattering  off spin
fluctuations is often overlooked since its magnitude is
substantially smaller than that of the Coulomb scattering. Due to
effective suppression of spin fluctuations in the ferromagnetic
phase, however, the temperature dependence of this relaxation
mechanism is much more pronounced. Indeed, in a fully spin polarized
state, the scattering takes place only off the quantum fluctuations
of localized spins. The total relaxation rate (\ref{tau}), which is
the sum of both contributions, restores its 20\% drop during
transition from paramagnetic to ferromagnetic phase. The majority of
this drop is found to be due to the suppression of the scattering
off localized spin fluctuations.

\begin{figure}
\centering
\includegraphics[width=0.9\linewidth]{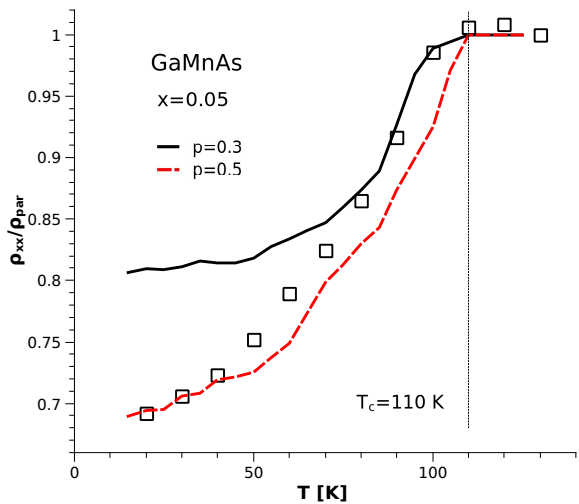}
\caption{Static resistivity of $\rm Ga_{0.95}Mn_{0.05} As$
normalized to the paramagnetic state value. Solid and dashed lines:
compensation levels $p=0.3$ and $p=0.5$ (number of holes per $\rm
Mn_{Ga}$). The vertical line indicates $T_c=110$ K. Experimental
points are from Ref.~\onlinecite{potashnik}.} \label{fig2}
\end{figure}

In Fig.~\ref{fig2} we plot the temperature dependence of the
static resistivity of Ga$_{0.95}$Mn$_{0.05}$As, normalized to the
paramagnetic state value. The calculation was done according to
Eqs.~(\ref{stat_cond}) and (\ref{tau})-(\ref{taus}). Solid and
dashed lines correspond to different levels of compensation in the
system, 0.3 and 0.5 hole per substitutional Mn, respectively (in
practice, this number is difficult to control). The open squares
represent the experimental data of Ref.~\onlinecite{potashnik}.
The theory demonstrates good agreement with experiment.

In summary, we have developed a theory of transport in spin and
charge disordered media within the valence band picture of
metallic GaMnAs. The approach combines the multiband ${\bf k\cdot
p}$ description of the semiconductor band structure with a
microscopic treatment of disorder and dynamical electron-electron
interaction by the methods of the memory-function formalism and
TDDFT. We applied our formalism to describe the experimentally
observed drop in static resistivity of GaMnAs in the ferromagnetic
phase.

This problem had been addressed before in Ref.~\onlinecite{lopez},
but with a model that was lacking some important features such as
scattering off the fluctuations of localized spins and
electron-electron interactions beyond RPA. Similar to
Ref.~\onlinecite{lopez}, we obtained agreement with the experimental
observations, but the underlying physics is quite different. Much of
the drop of resistivity in the ferromagnetic phase is found to be
due to the suppression of localized spin fluctuations in the
magnetically ordered state.

To conclude, we have developed a theoretical description of
itinerant carriers in DMSs within the valence band picture that
accounts for band structure, scattering from Coulomb and magnetic
impurities, and screening via dynamical many-body effects. An
accurate description of static transport properties in GaMnAs
involves a subtle interplay of all these ingredients. Our approach
is also suitable for frequency-dependent properties such as the
optical conductivity, as will be discussed elsewhere.

This work was supported by DOE Grant No. DE-FG02-05ER46213.

\end{document}